\documentclass{Interspeech2024}

\usepackage{multirow}
\usepackage{tabularx}
\usepackage{array}
\usepackage{xcolor}
\usepackage{makecell}
\usepackage{amssymb}
\usepackage{pifont}
\usepackage{cite}
\usepackage{hyperref}

\interspeechcameraready

\title{MaLa-ASR: Multimedia-Assisted LLM-Based ASR}

\name[affiliation={1}]{Guanrou}{Yang}
\name[affiliation={1}]{Ziyang}{Ma}
\name[affiliation={2}]{Fan}{Yu}
\name[affiliation={2}]{Zhifu}{Gao}
\name[affiliation={2}]{Shiliang}{Zhang}
\name[affiliation={1,\dagger}]{Xie}{Chen}

  \address{
  $^1$MoE Key Lab of Artificial Intelligence, AI Institute, X-LANCE Lab, \\ Shanghai Jiao Tong University, China,
  $^2$Alibaba Group, China}
\email{\{yangguanrou, chenxie95\}@sjtu.edu.cn}


\newcommand\blfootnote[1]{
  \begingroup
  \renewcommand\thefootnote{}\footnote{#1}
  \addtocounter{footnote}{-1}
  \endgroup
}

\begin{document}

\maketitle

\begin{abstract}

As more and more information-rich data like video become available, utilizing multi-modal auxiliary information to enhance audio tasks has sparked widespread research interest. The recent surge in research on LLM-based audio models provides fresh perspectives for tackling audio tasks. Given that LLM can flexibly ingest multiple inputs, we propose MaLa-ASR, an LLM-based ASR model that can integrate textual keywords extracted from presentation slides to improve recognition of conference content. MaLa-ASR yields average WERs of 9.4\% and 11.7\% on the L95 and S95 subsets of the SlideSpeech corpus, representing a significant relative WER drop of 27.9\% and 44.7\% over the baseline model reported in SlideSpeech. MaLa-ASR underscores LLM's strong performance in speech tasks and the capability to integrate auxiliary information conveniently. By adding keywords to the input prompt, the biased word error rate (B-WER) reduces relatively by 46.0\% and 44.2\%, establishing a new SOTA on this dataset.\footnote{Code and checkpoints are available at 
 https://github.com/X-LANCE/SLAM-LLM/tree/main/examples/mala\_asr\_slidespeech}

\end{abstract}
\blfootnote{$\dagger$ Corresponding author}
\vspace{-3mm} 

\section{Introduction}
Currently, vast amounts of data from various modalities are available online. Tasks involving multi-modal information have been widely studied, such as using visual and textual information from videos to aid in automatic speech recognition (ASR). Traditional multi-modal speech recognition models employ dedicated encoders to extract features from each modality and intricately design sophisticated architectures to effectively combine features from other modalities with audio features~\cite{xu2020discriminative,makino2019recurrent,yu2023hourglass, wang2024mlca}. The recent trend of large language model (LLM) based speech recognition models~\cite{wang2023lauragpt, wu2023decoder, fathullah2024prompting, yu2023connecting, ma2024embarrassingly, wu2023next, chen2024salm} offers novel perspectives to handle this task.

The profound understanding, generating, and generalization abilities of large language models acquired from extensive training data have resulted in ground-breaking performance across a wide range of text-based tasks. Recent work has found that large language models possess capabilities that extend beyond the text modality; they can understand information from other modalities such as images and audio.
Models like SALMONN~\cite{tang2023salmonn} and Qwen-Audio~\cite{chu2023qwen} are universal LLM-based audio models, capable of processing diverse audio types, including speech, sounds, and music, and adept at various audio tasks such as speech recognition, emotion recognition, and music caption. 

The conventional approach to employing an audio model with a large language model as its decoder starts by utilizing an encoder to extract features from the audio input. These features are then transformed into the text token space of the large language model through an adapter, commonly utilizing a linear projector or Q-former~\cite{BLIP2}. Subsequently, the processed audio embeddings are concatenated with text instruction prompts that are tailored for specific tasks and together fed into the large language model. Finally, the large language model generates the desired target text output autoregressively.

The input data for the large language model can be highly flexible and customized, provided that the inputs are compatible with the text token space of the LLM. Besides, current powerful large language models can typically accommodate extensive input sequences of several thousand tokens in length. Therefore, LLM-based audio models are inherently well-suited for integrating multi-modal auxiliary information and contextual information to aid in audio tasks.

In this paper, we propose an LLM-based speech model named MaLa-ASR that leverages multi-modal information to enhance the ASR task. A suitable dataset is SlideSpeech~\cite{wang2023slidespeech}, from which we put key textual information extracted from slides used during presentations into prompts to assist the LLM decoder in recognizing named entities and indistinctly pronounced words. MaLa-ASR, trained respectively on the L95 (473 hours) and S95 (161 hours) datasets, achieves average WERs of 9.4\% and 11.7\%, demonstrating a significant relative WER reduction of 27.9\% and 44.7\% compared to the contextual ASR baselines from SlideSpeech. This also reaffirms the excellent performance of LLM-based speech model architectures in speech tasks. By integrating keyword information, MaLa-ASR attains average WERs of 9.0\% and 11.2\%, marking relative WER reductions of 3.6\% and 4.1\% and significant B-WER reductions of 46.0\% and 44.2\% over the model without keywords. In addition, we investigate incorporating historical long-term contextual information into the prompts to enhance the speech recognition task; interestingly, this straightforward approach doesn't yield significant gains, which needs further exploration.

\vspace{-3mm} 
\section{Related works}
Humans often achieve more accurate understanding and judgment by processing multi-modal information. Multi-modal information can complement each other, eliminate ambiguities, and enhance comprehension~\cite{Hearinglips,Visualcontribution}. Recent research~\cite{ma2023auto,shi2022learning,haliassos2022jointly,ma2021end,sun2022tree} has increasingly endeavored to harness information from additional modalities to enhance audio tasks, with audio-visual speech recognition standing out as a primary focus of these investigations.
A series of studies focus on leveraging lip movements to assist speech recognition, which can significantly enhance accuracy, especially in environments with substantial background noise. Models like AV-HuBERT~\cite{shi2022learning} and RAVEn~\cite{haliassos2022jointly} exhibit superior performance on this task. They typically employ dedicated audio and visual encoders to extract features from the audio waveform and cropped lip images respectively. These features are then fused in various ways before being fed into a combined decoder including a projection layer and a Transformer decoder for hybrid CTC/attention~\cite{watanabe2017hybrid} training. The LRS3~\cite{afouras2018lrs3} and LRS2~\cite{yu2020audio} datasets are commonly utilized benchmarks for this task, complemented by datasets such as VoxCeleb2~\cite{chung2018voxceleb2} and AVSpeech~\cite{ephrat2018looking} for self-supervised pre-training, all of which predominantly feature video clips centered on the facial area. 

Another series of studies primarily utilizes textual information extracted from videos and images. SlideSpeech builds a large-scale AVSR dataset on online conference videos, primarily consisting of presentations delivered with slides. Similarly, SlideAVSR~\cite{wang2024slideavsr} constructs a dataset using scientific paper explanation videos. The slides intended to facilitate understanding, not only summarize key information related to the content being discussed but also include technical terms and named entities. Both works capture the middle frame image of each video segment and feed them into text detection (TD) and optical character recognition (OCR) modules to extract the texts in the slides. SlideSpeech utilizes a contextual ASR baseline model, which consists of a cross-attention module that allows speech embedding to integrate useful information from the contextual phrase embedding. SlideAVSR sends recognized texts as prompt to Whisper large-v3 model~\cite{radford2023robust} for fine-tuning and inference.
LCB-NET~\cite{yu2023lcbnet} proposes an innovative long-context biasing network for utilizing long-context information in videos. 
Another related work \cite{lakomkin2023end} utilizes an in-house dataset derived from publicly available videos on Instagram and Facebook. They integrate the video title and description that encompass information on topics and named entities as supplementary external context to improve speech recognition. Specifically, they integrate textual contexts into the prompt and pass it alongside audio tokens into the 7B-parameter LLaMA~\cite{touvron2023llama1} language model fine-tuned with LoRA~\cite{hu2021lora} adapters, which then generates the recognized spoken text.

\begin{figure}[t]
  \centering
  \includegraphics[width=\linewidth]{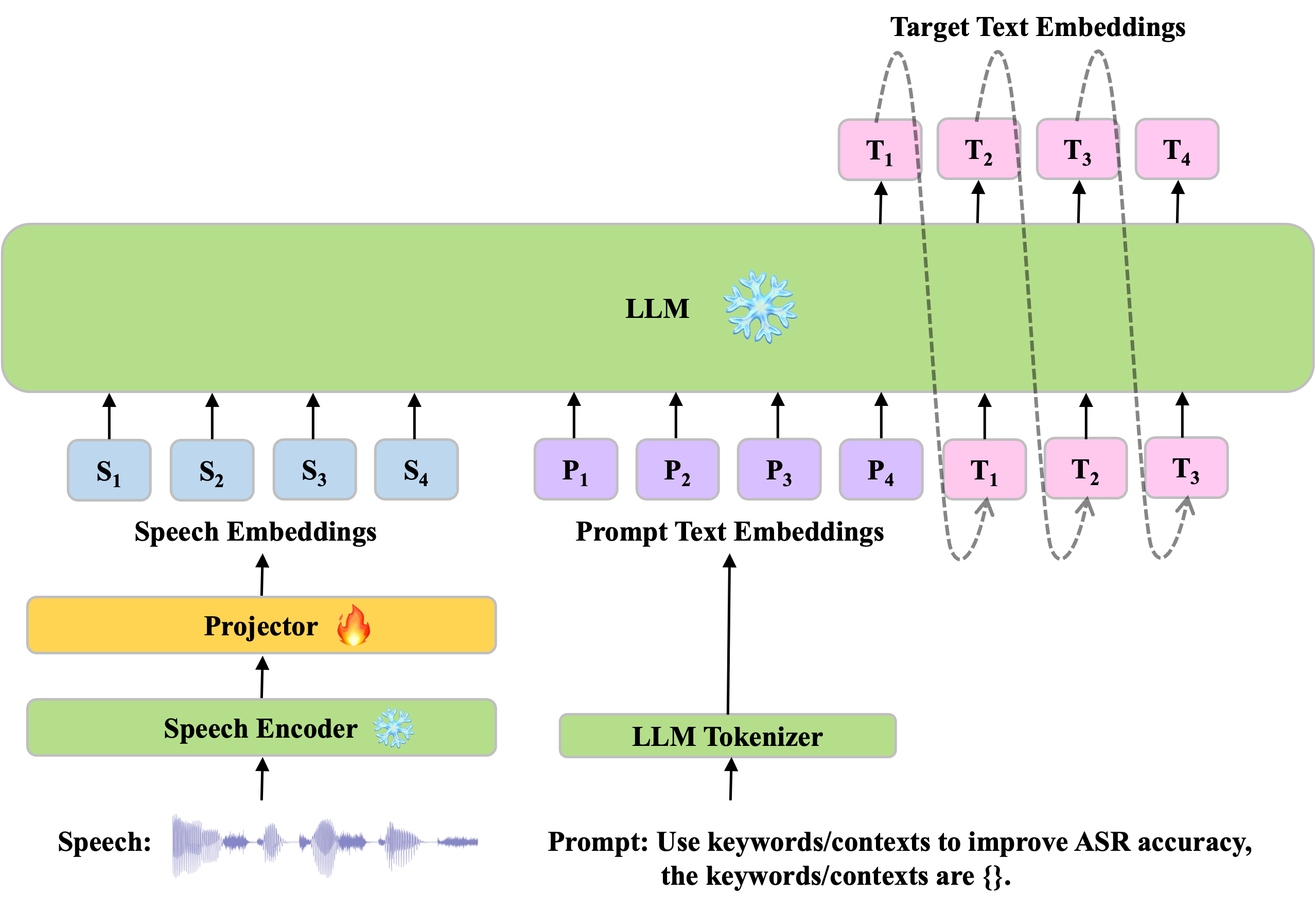}
  \caption{The model architecture of MaLa-ASR.  The main components are a frozen speech encoder, a frozen LLM, and a trainable projector. }
  \label{fig:model}
  \vspace{-7mm}
\end{figure}

\begin{table*}[ht]
  \caption{ASR results of MaLa-ASR with or without contextual keywords evaluated on dev/test datasets, trained on S95/L95 datasets.}
  \vspace{-2mm}
  \label{tab:keyword}
  \centering
  \resizebox{\linewidth}{!}{
  \begin{tabular}{llccccccccc}
    \toprule
    \multirow{2}{*}{\textbf{Train}} & 
    \multirow{2}{*}{\textbf{Model}} & 
     \multirow{2}{*}{\makecell{ \textbf{Contextual}\\ \textbf{Keywords}}} &
    \multicolumn{4}{c}{\textbf{Dev}} & \multicolumn{4}{c}{\textbf{Test}} \\
    \cmidrule(lr){4-7} \cmidrule(lr){8-11}
    & & & \textbf{WER} & \textbf{U-WER} & \textbf{B-WER} & \textbf{Recall $\uparrow$} & \textbf{WER} & \textbf{U-WER} & \textbf{B-WER} & \textbf{Recall $\uparrow$}  \\
    \midrule
    \multirow{5}{*}{\makecell{S95\\(161h)}} & 
    SlidesSpeech~\cite{wang2023slidespeech} & \ding{55}& 21.05 & 20.29&31.27&68.76 & 21.22 & 20.83&26.60&73.51 \\
     & MaLa-ASR & \ding{55}& 11.57 & \textbf{11.28} &16.23&83.83 & 11.80 & 11.71&13.52&86.71 \\ 
     & CPP~\cite{wang2023slidespeech} &\ding{51}& 20.80 & 20.22&28.61&71.48 & 20.95 & 20.73&24.05&76.10 \\
     & LCB-net~\cite{yu2023lcbnet} &\ding{51}& 18.80 & 18.11&27.90&72.09 & 19.21 & 18.89&23.70&76.48 \\
     & MaLa-ASR &\ding{51}&  \textbf{11.14} & 11.36&\textbf{8.92} & \textbf{91.44} & \textbf{11.26} & \textbf{11.52}& \textbf{7.67} & \textbf{92.50} \\
    \midrule
    \multirow{7}{*}{\makecell{L95\\(473h)}} & SlidesSpeech~\cite{wang2023slidespeech} & \ding{55}& 13.09 & 12.87&16.13&83.90 & 12.89 & 12.90&12.70&87.43 \\
     & MaLa-ASR (S1) &\ding{55}&   9.38 & 9.19&11.98&88.08 & 9.34& 9.33&9.52&90.64 \\ 
     & \qquad + LoRA & \ding{55}&  8.82 & 8.77&9.62&90.38 & 8.61& \textbf{8.72}&7.34&92.84 \\
     & CPP~\cite{wang2023slidespeech} & \ding{51}&  12.64 & 12.66&12.39&87.64 & 12.38 & 12.60&9.32&90.86\\ 
     & LCB-net~\cite{yu2023lcbnet} &\ding{51}& 12.21 & 12.21&12.12&87.98 & 12.02 & 12.24&9.03&91.12 \\
     & MaLa-ASR (S2) & \ding{51}& 8.91 & 9.13&6.07&94.02 & 9.14& 9.42&5.47&94.87 \\
     & \qquad + LoRA & \ding{51}& \textbf{8.30} & \textbf{8.53}& \textbf{5.22} & \textbf{94.87} & \textbf{8.46} & 8.73& \textbf{4.89} & \textbf{95.31} \\
    \bottomrule
  \end{tabular}
  }
\end{table*}

\begin{table*}[ht]
  \caption{Examples of recognition results on the SlideSpeech dataset. We present the corresponding slide, the ground truth transcription (GT), and the prediction result of MaLa-ASR with (AV) and without (A) contextual keywords. Incorrectly predicted words are marked in \textcolor{red}{red}. Words corrected using keyword information are highlighted in \textcolor{green}{green}. Words in slides but not corrected are marked in \textcolor{orange}{orange}.}
  \vspace{-2mm}
  \label{tab:example}
  \centering
  \resizebox{\linewidth}{!}{
  \begin{tabularx}{\textwidth}{lXX} 
    \toprule
    \textbf{Slide} & \includegraphics[width=\linewidth]{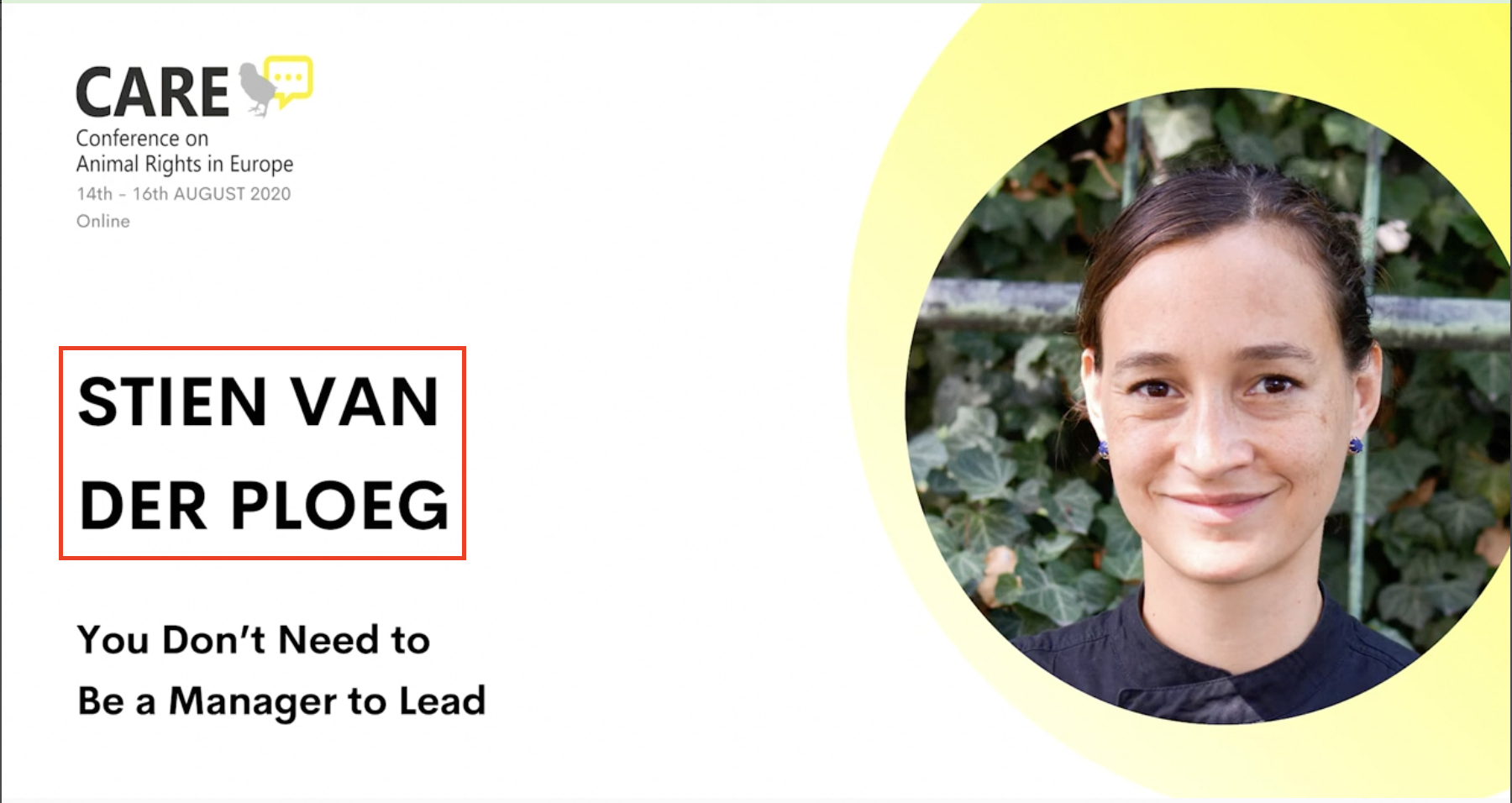} & \includegraphics[width=\linewidth]{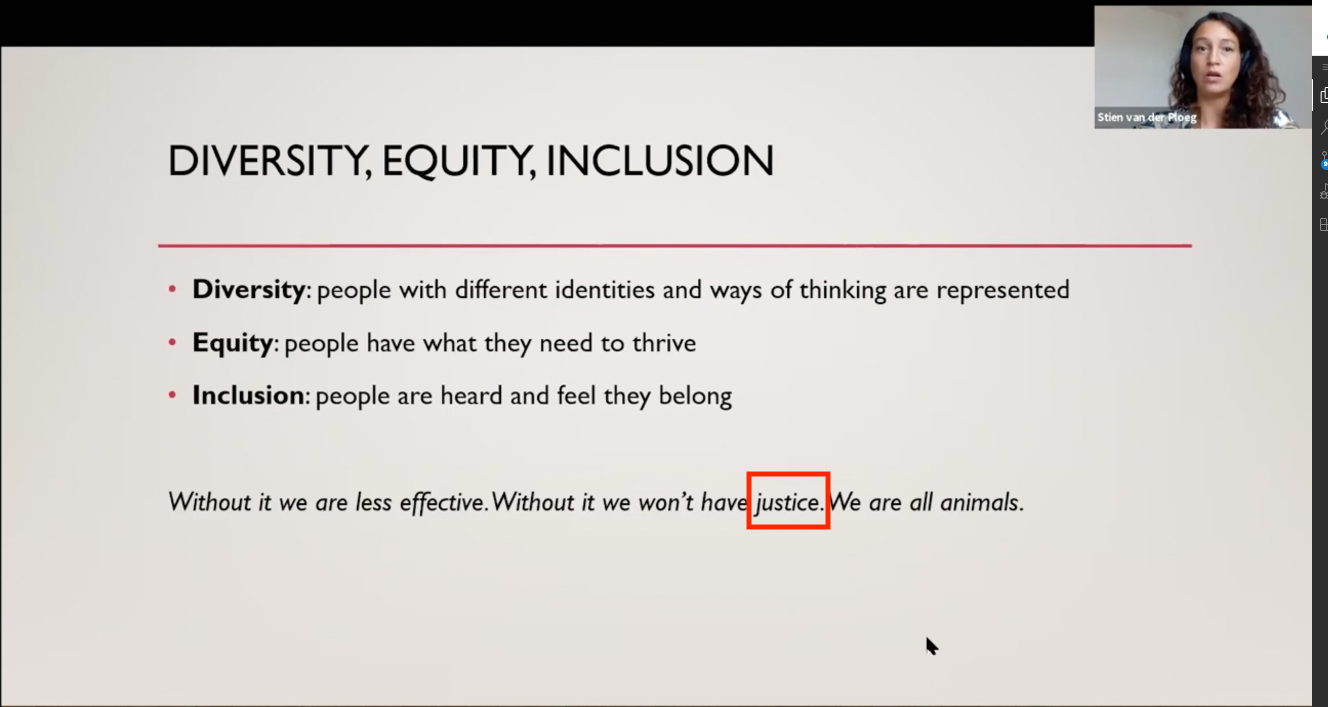} \\
    \midrule
    \textbf{GT} & okay so let me introduce our 1st speaker who is \textcolor{green}{stien van der ploeg} & without it we are less effective without it we will not have \textcolor{green}{justice} we are all animals  \\
    \midrule
    \textbf{AV} & okay so let me introduce our 1st speaker who is \textcolor{orange}{stein} \textcolor{green}{van der ploeg} & without it we are less effective without it we will not have \textcolor{green}{justice} we are all animals  \\
    \midrule
    \textbf{A} & okay so let me introduce our 1st speaker who is  \textcolor{orange}{steam} \textcolor{red}{funder plu} & without it we are less effective without it we will not have \textcolor{red}{guests} we are all animals  \\
    \bottomrule
  \end{tabularx}
  }
  \vspace{-5mm}
\end{table*}

\vspace{-3mm} 
\section{Proposed MaLa-ASR}
In this section, we will elaborate on the proposed MaLa-ASR, a multimedia-assisted LLM-based ASR model. As presented in Figure \ref{fig:model}, the overall model architecture consists of three parts, a speech encoder, a linear projector, and a large language model decoder. 
We use the official WavLM~\cite{chen2022wavlm} Large model as our speech encoder, which is pre-trained on the $94,000$ hours data including LibriLight~\cite{kahn2020libri}, GigaSpeech~\cite{chen2021gigaspeech} and VoxPopuli~\cite{wang2021voxpopuli}. WavLM is a powerful speech-based SSL model that excels at full-stack downstream speech tasks and is commonly used for extracting speech features. 
We use the public Vicuna 7B~\cite{chiang2023vicuna} as our large language model decoder. Vicuna is fine-tuned from the LLaMA base model by supervised instruction fine-tuning using conversations collected from ShareGPT. 
An adaptor is necessary for aligning the speech feature space with the LLM text token space. Inspired by SLAM-ASR~\cite{ma2024embarrassingly}, we use a simple-structured linear projector, consisting of a 1-D convolution layer that downsamples the 50Hz extracted speech features to 10Hz, and two linear layers with an intermediate hidden layer dimension of 2048.

The training process is as follows: The WavLM encoder accepts input speech sampled at 16kHz and outputs 50Hz feature sequences with a dimension of 1024. The feature sequences are then downsampled and projected, resulting in 10Hz speech embeddings with an expanded dimension of 4096. Prompts that contain task instructions and auxiliary information are processed through tokenization and encoding to produce prompt text embeddings. The speech and prompt text embeddings are concatenated to form a unified representation, which is then input into the LLM and decoded to yield the target transcripts autoregressively. We only train the lightweight projector and freeze the rest of the model. We calculate cross-entropy loss exclusively between the target hypothesis and transcripts as common practice for training LLM-based speech models.

\vspace{-2mm} 
\section{Experiments}
\subsection{Experimental Setup}

\vspace{-1mm} 
\subsubsection{Dataset}
We utilize the SlideSpeech~\cite{wang2023slidespeech} corpus as our dataset. The SlideSpeech corpus is a large-scale audio-visual dataset generated from online conference videos on YouTube, which has high-quality transcribed speech and synchronized slides. In addition to videos in 720p format and audio sampled at 16kHz, the corpus provides pre-processed OCR results and extracted keywords of the corresponding slides for each segment, enriching the dataset's utility. 

SlideSpeech has two training datasets of different sizes. The large training dataset (L95) includes 473 hours of audio, while the small training dataset (S95) contains 161 hours of audio, sampled from the large one. The dev and test sets consist of 5.07 and 8.75 hours of audio, respectively.

\vspace{-2mm} 
\subsubsection{Training}
The format of a training data sample is ``\textit{\textless speech\textgreater\ 
 USER: \textless prompt\textgreater\ ASSISTANT: \textless transcription\textgreater}''. \textit{\textless speech\textgreater} refers to the aligned speech embedding, which has the same dimension of 4096 as the LLM. 
To guide the LLM to leverage the keywords of the related slides, we design the \textit{\textless prompt\textgreater} as ``\textit{Transcribe speech to text. Use keywords in PPT to improve speech recognition accuracy. But if the keywords are irrelevant, just ignore them. The keywords are \{\}}'' and fill it with the corresponding keywords. When there are no keywords present, the prompt is simplified to ``\textit{Transcribe speech to text.}''
\textit{\textless transcription\textgreater} refers to the ground-truth transcription of the speech.

In the training process, only the lightweight projector (15.7M) is trained, while the speech encoder (315.5M) and the LLM (6.7B) are frozen. The LM loss is only calculated on \textit{\textless transcription\textgreater}. The model is trained for 110k steps. The learning rate increases linearly from 0 to the peaking rate of 5e-5 in the first 1k steps and then decays linearly to zero during the remaining training time. We use AdamW~\cite{loshchilov2017decoupled} optimizer with $\beta=(0.9,0.999)$ and zero weight decay. We conduct our experiments on 4 80GB A800 GPUs and set the batch size to 6. For the experiments with LoRA adapters, we add LoRA modules to the key, query, value, and output projection layers of every self-attention module. We set the rank as 32, the alpha as 32, and the dropout as 0.05, resulting in 33.6M extra trainable parameters of LLM. 

\vspace{-2mm} 
\subsubsection{Inference}
The format of a test data sample is ``\textit{\textless speech\textgreater\ USER: \textless prompt\textgreater\ ASSISTANT: }''. With this preceding input, the LLM will generate the transcription autoregressively. We use the beam search algorithm and set the beam size to 4 for decoding.

\vspace{-2mm} 
\subsection{Experimental Results}
\subsubsection{Results}
Table \ref{tab:keyword} presents the performance of MaLa-ASR on the SlideSpeech dataset. Following ~\cite{wang2023slidespeech}, we use WER, biased word error rate (B-WER), unbiased word error rate (U-WER), and the Recall of words in the keyword list to evaluate the performance. B-WER is calculated solely for words in the keyword list, while U-WER is computed for words not in the keyword list. Recall is the proportion of correctly recognized words that appear in both keyword lists and the ground truth transcription.

Our MaLa-ASR baseline model, trained separately on the L95 and S95 datasets, achieves average WER of 9.4\% and 11.7\%, with a notable average relative reduction in WER of 27.9\% and 44.7\% compared to the contextual ASR baseline model proposed in SlideSpeech.
After incorporating supplementary keyword information, MaLa-ASR achieves average WERs of 9.0\% and 11.2\% on L95 and S95. The WER is relatively reduced by 3.6\% and 4.1\% compared to the model without keywords. 

Upon integrating keyword information, the B-WER of MaLa-ASR notably decreases from 10.8\% to 5.8\% and from 14.9\% to 8.3\% on L95 and S95. The Recall rates improve from 89.4\% to 94.4\% and from 85.3\% to 92.0\%. Overall, there is a substantial reduction in B-WER and a clear increase in Recall, while the U-WER remained largely unchanged. This is visual proof of the model's effective utilization of keyword information related to the speech content from the prompts. Our model not only exhibits greater improvements in B-WER and Recall than the current existing Hotword models but also has a much simpler structure. The SlideSpeech baseline model employs a context encoder composed of a bidirectional LSTM, a multi-head cross-attention-based biasing layer, and a CPP network to integrate keywords. In contrast, we simply need to add the textual keywords to the prompt.


To further boost model performance, we fine-tuned the LLM with the LoRA adapter. With LoRA adapters, MaLa-ASR trained on L95 achieves average WERs of 8.7\%, an average relative decrease of 6.9\% in WER compared to the original MaLa-ASR.
Utilizing keyword information, the WER can be further reduced to 8.4\%.

Table \ref{tab:example} provides specific examples of MaLa-ASR correcting recognition results using keywords, particularly of named entities, proper nouns, and words that are pronounced unclearly.

\vspace{-2mm} 
\subsubsection{Analysis}

To gain deeper insights into how the model utilizes keywords, we specifically count how many keywords play a role in correcting the predictions. As shown in Table \ref{tab:correction}, approximately 88.8\% of the keywords are common words that are easily recognizable. MaLa-ASR is capable of identifying them correctly regardless of their presence in the prompt, which explains the seemingly modest average relative WER reduction of 3.9\% brought by keyword information. 53.3\% of the remaining keywords successfully helped the model correctly recognize them on the evaluation sets, demonstrating the effectiveness of the proposed method.

Considering the potential lack of auxiliary information in real-world test data, our ablation experiments in Table \ref{tab:ablation} reveal that the WER result of Model S3, trained with keywords but inferred without, is comparable to that of Model S1, which doesn't utilize keywords in both training and inference phases. This demonstrates the model's robustness and suggests its practicality in real-life settings where keywords are unavailable. The results from Model S4 demonstrate that if a model has not been trained to leverage keyword information, it cannot inherently utilize keywords based on prompts effectively.

\vspace{-1mm} 
\begin{table}[th]
\centering
\caption{The Proportion of keywords present in the transcriptions of the dev/test sets in four categories: correctly recognized by S1 and S2 model, only by S1 model, only by S2 model, and unrecognized by both models. S1/S2 refers to MaLa-ASR model trained without/with textual keywords on L95 dataset. \ding{51} and \ding{55} represents correct and incorrect recognition.
}
\label{tab:correction}
\begin{tabular}{lccc}
\toprule
\multicolumn{2}{c}{\multirow{2}{*}{Dev/Test}} & \multicolumn{2}{c}{S2} \\
\cmidrule{3-4}
& & \ding{51} &  \ding{55} \\
\midrule
\multirow{2}{*}{S1} & \ding{51}  & 87.3\% / 90.3\% & 1.1\% / 0.9\% \\
 & \ding{55} & 7.0\% / 5.0\% & 4.6\% / 3.8\% \\
\bottomrule
\end{tabular}
\vspace{-3mm}
\end{table}


\begin{table}[h!]
\centering
\caption{ASR results of proposed MaLa-ASR with or without contextual keywords during training and inference evaluated on dev/test datasets trained on L95 dataset.}
\label{tab:ablation}
\begin{tabular}{ccccc} 
\toprule
\multirow{2}{*}{Model} & \multicolumn{2}{c}{Contextual Keywords} & \multicolumn{2}{c}{WER} \\
\cmidrule(lr){2-3} \cmidrule(lr){4-5}
& Train & Infer & Dev & Test \\
\midrule
S1 & \ding{55} & \ding{55} & 9.38 & 9.34 \\
S2 & \ding{51} & \ding{51} & 8.91 & 9.14 \\
S3 & \ding{51} & \ding{55} & 9.53 & 9.47 \\
S4 & \ding{55} & \ding{51} & 9.81 & 9.64 \\

\bottomrule
\end{tabular}
\vspace{-5mm}
\end{table}
\vspace{-3mm} 
\subsection{Exploration of Utilizing Prior Context }

Common ASR models\cite{li2022recent} are trained and evaluated at the utterance level, however, in practical scenarios such as conversations and presentations, speech often presents in long-content forms. Since the content and logic of contexts are often closely linked, and certain words may be mentioned again, many studies attempt to integrate historical context information to better decode the current sentence. For instance, \cite{cui2023towards,gong2024advanced,chang2021context} investigate integrating historical context information in transducer architecture and \cite{kim2018dialog,hori2020transformer} model long-context scenarios in AED architecture.

Since MaLa-ASR can easily input auxiliary information, we also explore incorporating historical context information into the prompt to better assist the LLM in decoding the current speech.
Specifically, we intercept a certain length of transcribed text from the few sentences preceding the current one and fill it into the following prompt template: ``\textit{Using previous context:\{\}, improve speech recognition for this audio. Apply relevant details from the previous context.}'' As shown in Table \ref{tab:history}, with or without LoRA fine-tuning, historical context doesn't bring a significant gain, with the WER remaining approximately the same as the baseline. We design a variety of prompts and experiment with different context lengths, but interestingly, this method doesn't yield a substantial performance improvement. We hypothesize that certain keywords are closely and directly linked to the speech content, whereas the preceding information primarily aligns with the background, theme, and domain of the current sentence. Consequently, employing the same method to leverage preceding information proves ineffective. How LLM-based ASR models can utilize the historical context remains an intriguing question that warrants further exploration.

\vspace{-1mm} 
\begin{table}[h]
\centering
\caption{ASR results of proposed MaLa-ASR with or without historical contexts or LoRA fine-tuning evaluated on dev/test datasets trained on L95 dataset.}
\label{tab:history}
\begin{tabular}{lcccc}
\toprule
Model & History Context & Dev & Test \\
\midrule
MaLa-ASR  & \ding{55} & 9.38 & 9.34 \\
\qquad + LoRA  & \ding{55}  & 8.82 & 8.61 \\
MaLa-ASR  & \ding{51} & 9.30 & 9.38 \\
\qquad + LoRA  & \ding{51} & 8.74 & 8.72 \\ 
\bottomrule
\end{tabular}
\vspace{-3mm}
\end{table}

\vspace{-3mm} 
\section{Conclusion}
In this work, we make an effort to utilize multi-modal auxiliary information for improving ASR tasks with LLM-based speech model architecture. 
Our experiments indicate that the LLM-based speech model performs well, showcasing its substantial potential for speech tasks.
Additionally, it can effortlessly embed auxiliary information and effectively leverage it to enhance speech recognition, far surpassing the baseline model and establishing a new SOTA on the SlideSpeech dataset.
Besides, our straightforward method of incorporating historical long-term context into the prompt is not effective and requires further research. 
Our approach has certain limitations. The large language model with an extensive number of parameters demands considerable memory utilization and the autoregressive decoding method leads to reduced decoding speeds.
In the future, we will continue to explore improved methods of leveraging keywords and other approaches to effectively utilize slides.
We are also interested in using visual encoders to extract features from the slides, aiming at capturing useful structural information contained in images and text to assist in speech recognition.


\section{Acknowledgements}
\vspace{-0.15cm}
This work was supported by the National Natural Science Foundation of China  (No. 62206171 and No. U23B2018), Shanghai Municipal Science and Technology Major Project under Grant 2021SHZDZX0102, the International Cooperation Project of PCL and Alibaba Innovative Research Program.

\bibliographystyle{IEEEtran}
\bibliography{mybib}

\begin{thebibliography}{10}
\providecommand{\url}[1]{#1}
\csname url@samestyle\endcsname
\providecommand{\newblock}{\relax}
\providecommand{\bibinfo}[2]{#2}
\providecommand{\BIBentrySTDinterwordspacing}{\spaceskip=0pt\relax}
\providecommand{\BIBentryALTinterwordstretchfactor}{4}
\providecommand{\BIBentryALTinterwordspacing}{\spaceskip=\fontdimen2\font plus
\BIBentryALTinterwordstretchfactor\fontdimen3\font minus \fontdimen4\font\relax}
\providecommand{\BIBforeignlanguage}[2]{{%
\expandafter\ifx\csname l@#1\endcsname\relax
\typeout{** WARNING: IEEEtran.bst: No hyphenation pattern has been}%
\typeout{** loaded for the language `#1'. Using the pattern for}%
\typeout{** the default language instead.}%
\else
\language=\csname l@#1\endcsname
\fi
#2}}
\providecommand{\BIBdecl}{\relax}
\BIBdecl

\bibitem{xu2020discriminative}
B.~Xu, C.~Lu, Y.~Guo, and J.~Wang, ``Discriminative multi-modality speech recognition,'' in \emph{Proc. CVPR}, 2020.

\bibitem{makino2019recurrent}
T.~Makino, H.~Liao, Y.~Assael, B.~Shillingford, B.~Garcia, O.~Braga, and O.~Siohan, ``Recurrent neural network transducer for audio-visual speech recognition,'' in \emph{Proc. ASRU}, 2019.

\bibitem{yu2023hourglass}
F.~Yu, H.~Wang, Z.~Ma, and S.~Zhang, ``Hourglass-{AVSR}: Down-up sampling-based computational efficiency model for audio-visual speech recognition,'' in \emph{Proc. ICASSP}, 2024.

\bibitem{wang2024mlca}
H.~Wang, P.~Guo, P.~Zhou, and L.~Xie, ``{MLCA-AVSR}: Multi-layer cross attention fusion based audio-visual speech recognition,'' \emph{arXiv preprint}, 2024.

\bibitem{wang2023lauragpt}
J.~Wang, Z.~Du, Q.~Chen \emph{et~al.}, ``{LauraGPT}: Listen, attend, understand, and regenerate audio with {GPT},'' in \emph{arXiv preprint}, 2023.

\bibitem{wu2023decoder}
J.~Wu, Y.~Gaur, Z.~Chen, L.~Zhou, Y.~Zhu, T.~Wang, J.~Li, S.~Liu, B.~Ren, L.~Liu \emph{et~al.}, ``On decoder-only architecture for speech-to-text and large language model integration,'' in \emph{Proc. ASRU}, 2023.

\bibitem{fathullah2024prompting}
Y.~Fathullah, C.~Wu, E.~Lakomkin \emph{et~al.}, ``Prompting large language models with speech recognition abilities,'' in \emph{Proc. ICASSP}, 2024.

\bibitem{yu2023connecting}
W.~Yu, C.~Tang, G.~Sun, X.~Chen, T.~Tan, W.~Li, L.~Lu, Z.~Ma, and C.~Zhang, ``Connecting speech encoder and large language model for {ASR},'' in \emph{Proc. ICASSP}, 2024.

\bibitem{ma2024embarrassingly}
Z.~Ma, G.~Yang, Y.~Yang, Z.~Gao, J.~Wang, Z.~Du, F.~Yu, Q.~Chen, S.~Zheng, S.~Zhang \emph{et~al.}, ``An embarrassingly simple approach for llm with strong asr capacity,'' \emph{arXiv preprint}, 2024.

\bibitem{wu2023next}
S.~Wu, H.~Fei, L.~Qu, W.~Ji, and T.-S. Chua, ``Next-gpt: Any-to-any multimodal llm,'' in \emph{arXiv preprint}, 2023.

\bibitem{chen2024salm}
Z.~Chen, H.~Huang, Andrusenko \emph{et~al.}, ``Salm: Speech-augmented language model with in-context learning for speech recognition and translation,'' in \emph{Proc. ICASSP}, 2024.

\bibitem{tang2023salmonn}
C.~Tang, W.~Yu, G.~Sun, X.~Chen, T.~Tan, W.~Li, L.~Lu, Z.~Ma, and C.~Zhang, ``{SALMONN}: Towards generic hearing abilities for large language models,'' in \emph{Proc. ICLR}, 2024.

\bibitem{chu2023qwen}
Y.~Chu, J.~Xu, X.~Zhou \emph{et~al.}, ``{Qwen-Audio}: Advancing universal audio understanding via unified large-scale audio-language models,'' \emph{arXiv preprint}, 2023.

\bibitem{BLIP2}
J.~Li, D.~Li, S.~Savarese, and S.~Hoi, ``{BLIP-2}: Bootstrapping language-image pre-training with frozen image encoders and large language models,'' in \emph{Proc. ICML}, 2023.

\bibitem{wang2023slidespeech}
H.~Wang, F.~Yu, X.~Shi, Y.~Wang, and S.~Zhang, ``{SlideSpeech}: A large-scale slide-enriched audio-visual corpus,'' in \emph{Proc. ICASPP}, 2024.

\bibitem{Hearinglips}
H.~McGurk and J.~MacDonald, ``Hearing lips and seeing voices,'' \emph{Proc. Nature}, 1976.

\bibitem{Visualcontribution}
W.~H. Sumby and I.~Pollack, ``Visual contribution to speech intelligibility in noise,'' \emph{Proc. JASA}, 1954.

\bibitem{ma2023auto}
P.~Ma, A.~Haliassos, A.~Fernandez-Lopez, H.~Chen, S.~Petridis, and M.~Pantic, ``Auto-{AVSR}: Audio-visual speech recognition with automatic labels,'' in \emph{Proc. ICASSP}, 2023.

\bibitem{shi2022learning}
B.~Shi, W.-N. Hsu, K.~Lakhotia, and A.~Mohamed, ``Learning audio-visual speech representation by masked multimodal cluster prediction,'' \emph{arXiv preprint}, 2022.

\bibitem{haliassos2022jointly}
A.~Haliassos, P.~Ma, R.~Mira, S.~Petridis, and M.~Pantic, ``Jointly learning visual and auditory speech representations from raw data,'' \emph{arXiv preprint}, 2022.

\bibitem{ma2021end}
P.~Ma, S.~Petridis, and M.~Pantic, ``End-to-end audio-visual speech recognition with conformers,'' in \emph{Proc. ICASSP}, 2021.

\bibitem{sun2022tree}
G.~Sun, C.~Zhang, and P.~C. Woodland, ``Tree-constrained pointer generator with graph neural network encodings for contextual speech recognition,'' in \emph{Proc. Interspeech}, 2022.

\bibitem{watanabe2017hybrid}
S.~Watanabe, T.~Hori, S.~Kim \emph{et~al.}, ``Hybrid ctc/attention architecture for end-to-end speech recognition,'' \emph{Proc.JSTSP}, 2017.

\bibitem{afouras2018lrs3}
T.~Afouras, J.~S. Chung, and A.~Zisserman, ``{LRS3-TED}: a large-scale dataset for visual speech recognition,'' \emph{arXiv preprint}, 2018.

\bibitem{yu2020audio}
J.~Yu, S.-X. Zhang, J.~Wu \emph{et~al.}, ``Audio-visual recognition of overlapped speech for the {LRS2} dataset,'' in \emph{Proc. ICASSP}, 2020.

\bibitem{chung2018voxceleb2}
J.~S. Chung, A.~Nagrani, and A.~Zisserman, ``Voxceleb2: Deep speaker recognition,'' \emph{arXiv preprint}, 2018.

\bibitem{ephrat2018looking}
A.~Ephrat, I.~Mosseri, O.~Lang \emph{et~al.}, ``Looking to listen at the cocktail party: A speaker-independent audio-visual model for speech separation,'' \emph{arXiv preprint}, 2018.

\bibitem{wang2024slideavsr}
H.~Wang, S.~Kurita, S.~Shimizu, and D.~Kawahara, ``{SlideAVSR}: A dataset of paper explanation videos for audio-visual speech recognition,'' \emph{arXiv preprint}, 2024.

\bibitem{radford2023robust}
A.~Radford, J.~W. Kim, T.~Xu, G.~Brockman, C.~McLeavey, and I.~Sutskever, ``Robust speech recognition via large-scale weak supervision,'' in \emph{Proc. ICML}, 2023.

\bibitem{yu2023lcbnet}
F.~Yu, H.~Wang, X.~Shi, and S.~Zhang, ``{LCB-NET}: Long-context biasing for audio-visual speech recognition,'' in \emph{Proc. ICASPP}, 2024.

\bibitem{lakomkin2023end}
E.~Lakomkin, C.~Wu, Y.~Fathullah, O.~Kalinli, M.~L. Seltzer, and C.~Fuegen, ``End-to-end speech recognition contextualization with large language models,'' \emph{arXiv preprint}, 2023.

\bibitem{touvron2023llama1}
H.~Touvron, T.~Lavril, G.~Izacard, X.~Martinet, M.-A. Lachaux \emph{et~al.}, ``{LLaMA}: Open and efficient foundation language models,'' \emph{arXiv preprint}, 2023.

\bibitem{hu2021lora}
E.~J. Hu, Y.~Shen, P.~Wallis, Z.~Allen-Zhu, Y.~Li, S.~Wang, L.~Wang, and W.~Chen, ``Lora: Low-rank adaptation of large language models,'' \emph{arXiv preprint}, 2021.

\bibitem{chen2022wavlm}
S.~Chen, C.~Wang, Z.~Chen \emph{et~al.}, ``{WavLM}: Large-scale self-supervised pre-training for full stack speech processing,'' in \emph{Proc. JSTSP}, 2022.

\bibitem{kahn2020libri}
J.~Kahn, M.~Rivi{\`e}re, W.~Zheng, E.~Kharitonov, Q.~Xu, P.-E. Mazar{\'e}, J.~Karadayi, V.~Liptchinsky, R.~Collobert, C.~Fuegen \emph{et~al.}, ``Libri-light: A benchmark for asr with limited or no supervision,'' in \emph{Proc. ICASSP}, 2020.

\bibitem{chen2021gigaspeech}
G.~Chen, S.~Chai, G.~Wang \emph{et~al.}, ``Gigaspeech: An evolving, multi-domain {ASR} corpus with 10,000 hours of transcribed audio,'' in \emph{Proc. Interspeech}, 2021.

\bibitem{wang2021voxpopuli}
C.~Wang, M.~Riviere, A.~Lee \emph{et~al.}, ``Voxpopuli: A large-scale multilingual speech corpus for representation learning, semi-supervised learning and interpretation,'' in \emph{Proc. ACL}, 2021.

\bibitem{chiang2023vicuna}
W.-L. Chiang, Z.~Li, Z.~Lin, Y.~Sheng, Z.~Wu \emph{et~al.}, ``Vicuna: An open-source chatbot impressing {GPT}-4 with 90\%* {ChatGPT} quality,'' \emph{\url{https://vicuna. lmsys. org}}, 2023.

\bibitem{loshchilov2017decoupled}
I.~Loshchilov and F.~Hutter, ``Decoupled weight decay regularization,'' in \emph{Proc. ICLR}, 2019.

\bibitem{li2022recent}
J.~Li \emph{et~al.}, ``Recent advances in end-to-end automatic speech recognition,'' \emph{Proc. APSIPA}, 2022.

\bibitem{cui2023towards}
M.~Cui, J.~Kang, J.~Deng \emph{et~al.}, ``Towards effective and compact contextual representation for conformer transducer speech recognition systems,'' \emph{arXiv preprint}, 2023.

\bibitem{gong2024advanced}
X.~Gong, Y.~Wu, J.~Li, S.~Liu, R.~Zhao, X.~Chen, and Y.~Qian, ``Advanced long-content speech recognition with factorized neural transducer,'' \emph{Proc. TASLP}, 2024.

\bibitem{chang2021context}
F.-J. Chang, J.~Liu, M.~Radfar, A.~Mouchtaris, M.~Omologo, A.~Rastrow, and S.~Kunzmann, ``Context-aware transformer transducer for speech recognition,'' in \emph{Proc. ASRU}, 2021.

\bibitem{kim2018dialog}
S.~Kim and F.~Metze, ``Dialog-context aware end-to-end speech recognition,'' in \emph{Proc. SLT}, 2018.

\bibitem{hori2020transformer}
T.~Hori, N.~Moritz, C.~Hori, and J.~Le~Roux, ``Transformer-based long-context end-to-end speech recognition,'' in \emph{Proc. Interspeech}, 2020.

\end{thebibliography}

\end{document}